\documentclass[prd,preprintnumbers,nofootinbib,aps,twocolumn]{revtex4-1}
\usepackage{epsfig,amsmath,amssymb}
\usepackage{mathrsfs}
\usepackage[normalem]{ulem}
\usepackage[usenames,dvipsnames]{color}
\usepackage[pagebackref=false, colorlinks=false]{hyperref}

\begin{document}
\title{A Possible Quantum Gravity Hint in Binary Black Hole Merger}

\author{Parthasarathi Majumdar}
\email{bhpartha@gmail.com}
\affiliation{School of Physical Sciences, Indian Association for the Cultivation of Science, Kolkata 700032, India.} 

\begin{abstract}

We present a semi-rigorous justification of Bekenstein's Generalized Second Law of Thermodynamics applicable to a universe with black holes present, based on a generic quantum gravity formulation of a black hole spacetime, where the bulk Hamiltonian constraint plays a central role. Specializing to Loop Quantum Gravity, and considering the inspiral and post-ringdown stages of binary black hole merger into a remnant black hole, we show that the Generalized Second Law implies a lower bound on the non-perturbative LQG correction to the Bekenstein-Hawking area law for black hole entropy. This lower bound itself is expressed as a function of the Bekenstein-Hawking area formula for entropy. Results of the analyses of LIGO-VIRGO-KAGRA data recently performed to verify the Hawking Area Theorem for binary black hole merger are shown to be entirely consistent with this Loop Quantum Gravity-induced inequality. However, the consistency is independent of the magnitude of the LQG corrections to black hole entropy, depending only on the negative algebraic sign of the quantum correction. We argue that results of alternative quantum gravity computations of quantum black hole entropy, where the quantum entropy exceeds the Bekenstein-Hawking value, may not share this consistency.         
\end{abstract}

\maketitle

\section{Introduction}

It is a consensus view that GW150914 and subsequent similar observations by the LIGO consortium pertain to binary black hole (BBH) mergers to a black hole remnant \cite{LIGO1}-\cite{LIGO7}. To reinforce this standpoint, several research groups \cite{bad18}-\cite{bad22} have recently sought to investigate the validity of Hawking's theorem \cite{haw71} on the impossibility of decrease of the area of black hole horizons in any physical process, by more detailed analyses of the data on BBH coalescence. Recall that this theorem, as well as the other Laws of Black Hole Mechanics \cite{bch73} are based directly on classical general relativity, and as such, their verification from observational data is also an endorsement of that theory as the correct description of physical spacetime. 

Inspired by ref. \cite{bch73}, Bekenstein \cite{bek73} proposed that in a universe with black holes present, a Generalized Second law of Thermodynamics must hold, in which the entropy of black holes (which was supposed to originate from a quantum theory of gravity) is taken into account. Taken together with Bekenstein's other hypothesis that black hole entropy must be a (linear) function of the horizon area, and adopting confirmatory arguments from Hawking's seminal work on black hole radiance \cite{haw75}, these proposals are the key pillars on which Black Hole Thermodynamics is founded. The Generalized Second Law reduces to Hawking's area theorem when restricted to classical general relativity. Calculations of black hole entropy in LQG \cite{abck98} and in superstring theory (restricted to five dimensional extremal black holes) \cite{vafs96} both confirm the BH area law. However, following Bekenstein's argument that black hole entropy must have quantum gravity origins, one expects specific corrections to the classical area theorem for every serious proposal of quantum gravity. 

In this paper, we first attempt a semi-rigorous justification of both Bekenstein's hypotheses based on a generic formulation of quantum gravity where the role of the quantum Hamiltonian constraint is highlighted. We next specialize to Loop Quantum Gravity (LQG) where an ab initio non-perturbative computation of black hole entropy has been performed by different groups over two decades \cite{km98} - \cite{abhi-pm14}, leading to specific corrections to the semi-classical Bekenstein-Hawking (BH) entropy. These LQG corrections are themselves functions only of the BH entropy. Incorporating these corrections into the Generalized Second Law as applied to BBH coalescence studied for almost a decade by the LVK collaboration, an inequality emerges, giving an estimate of the magnitude of the LQG corrections. This inequality can be expressed directly in terms of the measured total horizon area (BH entropy) of the inspiralling black holes, very much prior to merger, and in terms of the `area (BH entropy) excess' deduced much later, post-ringdown, from the measured area of the merger remnant. The successful assay on verification of the Hawking area theorem \cite{bad18} - \cite{bad22} is then used to show that our LQG bound is entirely consistent with results of these analyses of LVK data.             

\section{Generalized Second Law}

\begin{figure}[h]
\begin{center}
\includegraphics[width=7cm]{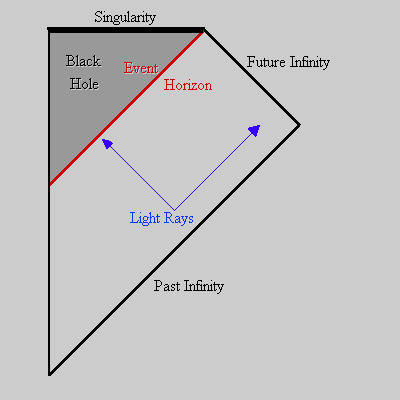}
\label{Fig.1}
\caption{Classical black hole spacetime}
\end{center}
\end{figure}

A generic classical black hole spacetime, depicted in Fig. 1, can be described mathematically by ${\cal B} = {\cal M} - {\cal J}^{-} ({\cal I}^+)$, where ${\cal M}$ is the entire spacetime and ${\cal J}^-({\cal I}^+)$ is the chronological past of asymptotic future null infinity. The {\it inner} boundary of ${\cal B}~,~\partial {\cal B}= h_+$ is called the future event horizon. 

The quantum description of such a spacetime may begin from the assumption of the Hilbert space of the system ${\cal H}$ having the structure ${\cal H}_{\cal B} \otimes {\cal H}_{h_+}$. Any general state $|\Psi \rangle \in {\cal H}$ can then be expanded as
\begin{eqnarray}
| \Psi \rangle = \sum_{{\cal B},h_+} C_{{\cal B} h_{+}} |\psi_{\cal B} \rangle \otimes |\psi_{h_+} \rangle  \label{gens}
\end{eqnarray} 
where, the complex matrix coefficients $C_{{\cal B} h_{+}}$ are not necessarily diagonal, thereby permitting possible entanglement between the bulk (${\cal B}$) and horizon (boundary) states. The Hamiltonian for the spacetime is assumed to have the structure ${\hat H} = {\hat H}_{\cal B} \otimes {\bf I}_{h_+} \oplus {\bf I}_{\cal B} \otimes {\hat H}_{h_+}$, i.e., ${\hat H}_{\cal B}$ acts only on $|\psi_{\cal B} \rangle \in {\cal H}_{\cal B}$, while ${\hat H}_{h_+}$ acts only on $| \psi_{h_+} \rangle \in {\cal H}_{h_+}$. A third and important assumption is that states of the black hole Hilbert space ${\cal H}_{\cal B}$ are {\it solutions} of the quantum Hamiltonian constraint : ${\hat H}_{\cal B} | \psi_{\cal B} \rangle = 0$. 

As a consequence, it follows that the `average energy' of the system
\begin{eqnarray}
\langle \Psi | {\hat H} | \Psi \rangle &=& \sum_{h_+} D_{h_+} \langle \psi_{h_+} | {\hat H}_{h_+} |\psi_{h_+} \rangle \nonumber \\
D_{h_+} &=& \sum_{\cal B} |C_{{\cal B}, h_+}|^2 || |\psi_{\cal B} \rangle ||^2. \label{ave}
\end{eqnarray}
 
We now consider a canonical ensemble of such spacetimes in equilibrium with a heat bath with an inverse temperature $\beta$; the canonical partition function is given by the standard definition : ${\cal Z} = Tr \exp -\beta {\hat H}$ where the trace is over all states $|\Psi \rangle \in {\cal H}$. Eqn (\ref{ave}) is now seen to imply that ${\cal Z} = Tr_{h_+} \exp -\beta {\hat H}_{h_+} \equiv {\cal Z}_{h_+}(\beta)$. Thus the Hamiltonian constraint reduces the thermodynamics of the system to the thermodynamics of the horizon states which then serve as microstates for computation of the canonical entropy of the system. If these horizon states also diagonalize a suitably defined {\it area operator}, then the canonical entropy 
\begin{eqnarray}
S(\beta) \equiv \left( 1 + \frac{\partial}{\partial \log \beta} \right ) {\cal Z}_{h_+} = S(A_{h_+}) \label{are}
\end{eqnarray}
Thus, somewhat heuristically, we are led to Bekenstein's contention that black holes must have an entropy (gravitational in character) which is to be a function of the horizon area. Further, he hypothesized \cite{bek73} that the functional form of this entropy must be {\it linear}, which when reinforced by Hawking's seminal work on black hole radiance \cite{haw75}, leads to the Bekenstein-Hawking area law for black hole entropy $S_{BH}(A_{h_+}) = A_{h_+} / 4 l_P^2$ with $l_P$ being the Planck length. The veracity of the area law has been verified in ab initio calculations in several serious proposals of quantum gravity, including loop quantum gravity \cite{abck98} (for four dimensional generic black holes), and for five dimensional extremal black holes in string theory \cite{vafs96}. In the former case however, quantum spacetime fluctuations \cite{km98} - \cite{abhi-pm14}  lead to a whole slew of quantum corrections to the Bekenstein-Hawking area law, as briefly recapitulated in the next section.

We end this section with the observation that if two black holes, initially far away, orbit each around other, leading to an eventual merger to a remnant black hole with emission of gravitational waves, treating this is as an isolated system, the thermodynamic second law of entropy increase would imply that 
\begin{eqnarray}
S_{bh}(A_{h_+}) + S_{GW} \geq S_{bh1}(A_{h_{1+}}) + S_{bh2}(A_{h_{2+}}). \label{g2l}
\end{eqnarray}
where, $S_{bh}$ is the entropy of the remnant black hole, while $S_{bh1}, S_{bh2}$ are entropies of the inspiralling ones. This is known as the Generalized Second Law of thermodynamics in a universe where black holes are present and may merge emitting gravitational waves. It is obvious that if $S_{bh} = S_{BH}$, then (\ref{g2l}) is just a simple addendum to Hawking's classical black hole area theorem \cite{haw71}. However, with quantum spacetime corrections to $S_{bh}$ beyond the area law, eqn (\ref{g2l}) may imply further non-trivial predictions. 

\section{Quantum spacetime corrections to Bekenstein-Hawking entropy} 

{\it Isolated} horizons \cite{abf00}-\cite{abl01}, a non-stationary generalization of stationary event horizons, are a particularly useful concept for the ab initio computation of black hole entropy. Classically, the symplectic structure on such horizons is that of an $SU(2)$ Chern-Simons theory of connections which are pullbacks of the spacetime connection in the first order formulation of general relativity, to the spherical foliation of the horizons. Solder two-forms constructed from bulk densitized triads in the Sen-Ashtekar formulation of general relativity represent sources for the horizon Chern-Simons connection fields. In bulk LQG \cite{ashlew}, holonomies of connections along the edges of spin network and the fluxes of the densitized triads through surfaces bounded by the edges, represent the quantum degrees of freedom. The inner boundary of this quantum geometrical structure is a punctured $S^2$ (for non-rotating isolated horizons), with punctures carrying spin deposited on the $S^2$ by bulk spin network edges. In this framework, fluxes are distributional, thus providing `pointlike' sources for the quantum Chern-Simons field strength. The states on the punctured $S^2$ are the microstates in a microcanonical ensemble, being the states of the $SU(2)$ Chern-Simons theory coupled to the spins at punctures \cite{km98}. 

\begin{figure}[h]
\begin{center}
\includegraphics[width=7cm]{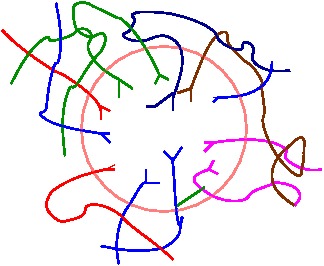}
\label{Fig.2}
\caption{Quantum black hole}
\end{center}
\end{figure}      

The dimensionality of the Hilbert space of these states is itself related to the {\it number} of conformal blocks of the conformally invariant $SU(2)_k$ Wess-Zumino-Witten model that exists on a spatial foliation of the isolated horizon with punctures at the location of the sources. For large $k$, this  number can be computed in terms of the spins \cite{km98}, yielding, for a spin configuration $j_1, ...j_P$
\begin{eqnarray}
{\cal N}(j_1, ...j_P)  &=& \prod_{i=1}^{P} \sum_{m_i=-j_i}^{j_i} [ \delta_{\sum_{n=1}^P m_n,0} \nonumber \\
&-& \frac12  \delta_{\sum_{n=1}^P m_n,-1} -\frac12 \delta_{\sum_{n=1}^P m_n,1} ] .\label{dimcs}
\end{eqnarray}
The total number of states is given by 
\begin{eqnarray}
{\cal N} = \sum_P \prod_{i=1}^P \sum_{j_i} {\cal N}(j_1, ...j_P).
\end{eqnarray}
Usual Boltzmann entropy is given by $S= \log {\cal N}$, and in the limit of large $k=A/l_P^2$, one obtains for the microcaonical entropy of quantum isolated horizons\cite{km2000}-\cite{abhi-pm14}, the result
\begin{eqnarray}
S_{bh} = S_{BH} - \frac32 \log S_{BH} +{\cal O}(S_{BH}^{-1}) ~, \label{qent}
\end{eqnarray}     
where, $S_{BH} \equiv A_{h+}/4l_P^2$ is the semiclassical Bekenstein-Hawking area law for any black hole with $A_{h+}$ being the cross-sectional area of the horizon, and $l_P = (G \hbar/c^3)^{1/2}$ is the Planck length. In some of the cited works it has been claimed that the isolated horizon states are those of a $U(1)$ Chern-Simons theory; however, as shown in ref. \cite{bkm09}, taking account of the additional gauge fixing in these papers, the corrections given in (\ref{qent}) remain valid.
\section{Prediction from the Generalized Second Law}

We define the remnant Bekenstein-Hawking entropy $S_{BH}(A_{h_+}) \equiv S_{BHr}$, and the inspiral black holes have $ S_{BH}(A_{h_{1+}}) \equiv S_{BH1}, S_{BH}(A_{h_{2+}}) \equiv S_{BH2}$, so that the Generalized Second Law (\ref{g2l}) can be re-expressed, including the LQG corrections in (\ref{qent}), as 
\begin{eqnarray}
S_{BHr} + S_{GW} -\frac32 \log S_{BHr} & \geq & S_{BH1} + S_{BH2} \nonumber \\
&-& \frac32 \log ( S_{BH1} S_{BH2})  \label{ine1}
\end{eqnarray}
Defining $S_{BHi} \equiv S_{BH1} + S_{BH2}$ as the inspiral black hole entropy, and $\Delta S_{BH} \equiv S_{BHr} - S_{BHi}$ as the change in entropy due to the coalescence, the inequality (\ref{ine1}) can be rewritten as 
\begin{eqnarray}
\Delta S_{BH} + S_{GW} \geq \log \left( \frac{S_{BH1} S_{BH2}}{S_{BHr}} \right)^{-3/2}. \label{ine2}
\end{eqnarray}
This can be reorganized and expressed in terms of direct measureables
\begin{eqnarray}
&& S_{BHi}^{-1}\log \left \{ \frac{ S_{BHi} [ 1 - (\delta_{12}S_{BH} / S_{BHi})^2]} {4[1 + (\Delta S_{BH}/ S_{BHi})]} \right \}   \geq \nonumber \\
& -& \frac23 \frac{\Delta S_{BH} + S_{GW}}{S_{BHi}} \label{ine3}   
\end{eqnarray} 
where $\delta_{12}S_{BH} \equiv |S_{BH1} - S_{BH2}|$. 

A perusal of the analyses in ref.s \cite{bad18}-\cite{bad22} reveals that the relative entropy excess $\Delta S_{BH}/S_{BHi} \in [\Delta_{max}S_{BH}/S_{BHi}~,~\Delta_{min}S_{BH}/S_{BHi}]$ where error bars have been taken into account. This enables rewriting (\ref{ine3}) as a strict inequality
\begin{eqnarray}
&& S_{BHi}^{-1}\log \left \{ \frac{ S_{BHi} [ 1 - (\delta_{12}S_{BH} / S_{BHi})^2]} {4[1 + (\Delta_{min} S_{BH}/ S_{BHi})]} \right \} > \nonumber \\
& -& \frac23 \frac{\Delta_{max} S_{BH} + S_{GW}}{S_{BHi}} \label{ine4}   
\end{eqnarray} 
We now make a few approximations : clearly for BBH mergers like GW150914, the inspiralling black holes are similar, such that $(\delta_{12} S_{BH}/S_{BHi})^2 << 1$; likewise, the analyzed data from ref.s \cite{bad18}-\cite{bad22} shows that $\Delta_{min} S_{BH}/S_{BHi} << 1$. As regards the gravitational wave entropy $S_{GW}$, a preliminary estimate made in ref. \cite{mr21} implies that $S_{GW}/S_{BHi} << \Delta_{max} S_{BH}/S_{BHi}$. With these approximations, the inequality (\ref{ine4}) reduces to 
\begin{eqnarray}
\frac{\log(S_{BHi}/4)}{S_{BHi}} > - \frac23 \frac {\Delta_{max} S_{BH}}{S_{BHi}} . \label{res}
\end{eqnarray}
That this inequality is valid is obvious from the data analyses of ref.s \cite{bad18}-\cite{bad22}: $S_{BHi} >>4$, ensuring that the {\it lhs} is strictly positive. Also, for most data $\Delta_{max}S_{BH}/S_{BHi} > 1$, rendering the {\it rhs} strictly negative. Thus, as mentioned earlier, the lower bound on LQG corrections to the BH entropy, derived by substitution in the Generalized Second Law, is entirely consistent with the analyzed LVK data. This is perhaps the first time that a prediction based on a quantum gravity proposal has been fully borne out by LVK data on BBH mergers. 

\section{Discussion}

We should perhaps emphasize that our claim of consistency of the outcome of combining the loop quantum gravity result for quantum black hole entropy and the Generalized Second Law, with the analyses in ref.s \cite{bad18}-\cite{bad22} of the LVK data on gravitational waves, is {\t not} a claim that LVK data has now given a precise confirmation of a quantum gravity result. The present observational accuracy of the data would have to increase many, many orders of magnitude before such a claim may be made. The main point of the paper is that the consistency pointed out in the paper in inequality (12), holds {\it irrespective} of the value of the constant prefactor $-2/3$ in the {\it rhs}, so long as it is a negative real number ! The number $-2/3$ emerges from the loop quantum gravity calculation of black hole entropy whose result has been presented in eqn (7), and is likely to differ for calculations of black hole entropy based on other approches to quantum gravity. All such results with a negative coefficient of the logarithmic correction to the area law are consistent with the analyses of LVK data which confirms Hawking's area theorem. A finer distinction between members of this class of theories of quantum gravity, i.e., where the quantum black hole entropy is {\it less than} the Bekenstein-Hawking value, is of course not possible with present levels of accuracy.  

Contrast this situation with those quantum gravity approaches in which the logarithmmic correction appears with a positive constant, i.e., $S_{bh} = S_{BH} + \xi \log S_{BH} + \cdots $ for macroscopic black holes, with a positive real $\xi$. The brief analysis performed above now leads to an upper bound on the correction to the area law, which transcribes into 
\begin{eqnarray*}
\frac{\log(S_{BHi}/4)}{S_{BHi}} < \xi^{-1} \frac {\Delta S_{BH}}{S_{BHi}}
\end{eqnarray*}    
Unlike in the case for loop quantum gravity-like calculations, both sides of this inequality are now positive numbers. The consistency with the aforementioned analyses of LVK data is now no longer guaranteed. For a range of values of $\xi$, a tension may arise between this theoretical result and the analyses of LVK data confirming the area theorem. 

To me, discerning this power of the data and its analyses to discriminate between two distinct classes of quantum gravity calculations of black hole entropy is {\it unexpected} and therefore {\it very novel}. What is remarkable is that the accuracy, with which the analyses presented in ref.s [9]-[11] claim the validation of Hawking's area theorem, is sufficient to demonstrate full consistency of LQG-like calculation of quantum black hole entropy with those analyses. No additional accuracy of the data is required for this demonstration.  Intuitively, for reasons which are intriguing but perhaps not completely transparent at this point, the area theorem analyses of LVK data seem to favour quantum black hole entropy calculations in which the quantum entropy is {\it less} than the Bekenstein-Hawking value, rather than those in which quantum corrected entropy exceeds that value.   

While observational accuracy in gravitational wave detection continues to increase with time, further discrimination on the basis of the data between quantum gravity results in the same category may gradually emerge in future. 

We should additionally mention the caveat that a key assumption regarding comparison with LVK data is that the inspiralling as well as post-merger remnant black holes in a BBH merger are slowly spinning so that the non-rotating approximation is approximately applicable. The LQG corrections to the Bekenstein-Hawking entropy constitute a robust result in the non-rotating regime. For rotating horizons, there are ambiguities in the LQG approach to the calculation of black hole entropy, which are yet to be satisfactorily resolved. It is hoped that once these issues are resolved, a similar consistency with LVK data, as we have sought to present here, will emerge. We hope to report on this in a future publication.

\section{Acknowledgements}

I thank Prof Badri Krishnan and Soumendra Kishore Roy for providing me with ref.s \cite{bad06} - \cite{bad22}

\end{document}